\documentclass[letterpaper, 10 pt, conference]{ieeeconf} 
\IEEEoverridecommandlockouts  
\usepackage{cite}
\usepackage{graphics}
\usepackage{graphicx}
\usepackage{epsfig}
\usepackage{multirow}
\usepackage{amssymb}
\usepackage{hyperref}
\usepackage{subcaption}
\usepackage{amsmath}
\usepackage{booktabs}
\usepackage{longtable}
\usepackage{stackengine}
\usepackage{xfrac}
\usepackage{tabularx}
\usepackage[dvipsnames]{xcolor,colortbl}
\usepackage[vlined,boxruled]{algorithm2e}  
\usepackage[clock]{ifsym} 
\usepackage{float} 
\usepackage[normalem]{ulem}

\usepackage{bm} 

\usepackage{soul}

\begin{document}

\graphicspath{{Figures/}}
\allowdisplaybreaks

\title{Polynomial Chaos-based Input Shaper Design under Time-Varying Uncertainty}
\author{Johannes G{\"u}ttler, Karan Baker, Premjit Saha, James Warner, and Adrian Stein
\thanks{J. G{\"u}ttler is with the Goethe-Universität Frankfurt am Main, 60629 Frankfurt am Main, Germany. {\tt\small(email: johannes.k.guettler@gmail.com)}.}
\thanks{K. Baker and A. Stein are with the Department of Mechanical and Industrial Engineering, Louisiana State University, LA 70803, USA. {\tt\small(email: \{kbake54,astein\}@lsu.edu)}.}
\thanks{Premjit Saha is with the University at Buffalo, Buffalo, NY 14260, USA. {\tt\small(email: premjits@buffalo.edu)}.}
\thanks{J. Warner is with the NASA Langley Research Center. {\tt\small(email: james.e.warner@nasa.gov)}.}}

\maketitle

\begin{abstract}

The work presented here investigates the application of polynomial chaos expansion toward input shaper design in order to maintain robustness in dynamical systems subject to uncertainty. Furthermore, this work intends to specifically address time-varying uncertainty by employing intrusive polynomial chaos expansion. The methodology presented is validated through numerical simulation of intrusive polynomial chaos expansion formulation applied to spring mass system experiencing time-varying uncertainty in the spring stiffness. The system also evaluates non-robust and robust input shapers through the framework in order to identify designs that minimize residual energy. Results indicate that vibration mitigation is achieved at a similar accuracy, yet at higher efficiency compared to a Monte Carlo framework. 

\end{abstract}

\IEEEpeerreviewmaketitle

\vspace{-0.1in}
\section{INTRODUCTION}
Uncertainty quantification (UQ) is vital for designing reliable and robust engineering systems, particularly when system parameters are subject to bounded intervals or evolve dynamically over time \cite{wang_polynomial_2020, wang_polynomial_2024, xia_dynamic_2016, yang_reliability-constrained_2025, yang_uncertain_2024}. Classical probabilistic approaches often assume fixed distributions and time-invariant uncertainties, limiting their applicability for complex real-world scenarios such as Unmanned Aerial Vehicle-payload control, spacecraft attitude tracking, and structural dynamics \cite{ni_using_2019, hu_uncertainty_2016, sun_pce-based_2022}. Polynomial Chaos Expansion (PCE) provides an efficient spectral framework to represent stochastic system responses via orthogonal polynomial basis functions, enabling fast uncertainty propagation and sensitivity analysis \cite{sudret_global_2008, stein_shapley_2023, singh_polynomial_2010}. Non-intrusive PCE approaches are widely used, but intrusive formulations, which embed uncertainty expansions directly into system equations, offer improved accuracy and computational benefits for dynamical systems \cite{harman_applying_2016, singh_polynomial_2010}. Arbitrary PCEs have been developed, requiring only finite moments or raw data and allowing flexible data-driven UQ \cite{oladyshkin_data-driven_2012}. Hybrid methods combining PCE with set-based techniques such as zonotopic and interval analysis reduce conservatism and enhance interval estimation accuracy, particularly for nonlinear and fuzzy systems with time-invariant parameters \cite{wang_polynomial_2024, wang_interval_2021, wang_polynomial_2020}. Further, second-order uncertainty models and scalable probabilistic frameworks using, for instance, Johnson distributions enable unified treatment of aleatory and epistemic uncertainties without relying on independence or exact distributional forms \cite{utkin_second-order_2003, zaman_probabilistic_2011}. Time-dependent interval process models and dynamic response analyses have also been introduced to capture evolving uncertainties in structures and control systems, where sliding mode control and iterative optimal attitude control methods can be applied for vibration control \cite{xia_dynamic_2016, yang_uncertain_2024, yang_reliability-constrained_2025}. 

Despite these advances, it has been shown that classical PCE methods degrade in long-time integrations because the fixed orthogonal polynomial basis becomes suboptimal as the probability density functions (pdf) of the system response evolve, leading to a loss of accuracy and convergence; this motivates the development of adaptive or time-dependent PCE approaches. Time-dependent generalized polynomial chaos methods can adapt polynomial bases over time to maintain accuracy \cite{gerritsma_time-dependent_2010}. Surrogate models using sparse PCE combined with stochastic time warping have also improved oscillatory system analyses in the past\cite{mai_surrogate_2017}.

In recent years, robust input shaping and sensitivity analyses that leverage global sensitivity metrics can facilitate uncertainty-aware control design in vibration control \cite{stein_global_2022, stein_shapley_2023, stein_shapley_2025, singh_optimal_2009}. Time-delay filters (TDF) are the implementation framework of input shapers, where the shaper’s sequence of impulses is realized as delayed and weighted inputs to cancel system vibrations. Experimental applications include real-time input shaping for gantry cranes and minimum-time input shaping under velocity constraints \cite{stein_aruco_2024, stein_minimum_2023}. Moreover, adaptive truncated PCE algorithms have been proposed for robust trajectory optimization in ascent launch vehicles facing nonlinear, fast time-varying uncertainties \cite{sun_pce-based_2022}. Building upon these foundations, this study proposes an intrusive PCE framework with time-dependent uncertainty in dynamical systems. The example showcased here is an undamped spring-mass system with uncertain spring stiffness over time.

The novelty of this work is to design precision motion controllers under influence of time-varying uncertainties with global sensitivity analysis (GSA). By adapting the polynomial basis dynamically at specific time instants, this method enhances vibration suppression and robust motion control. The approach combines theoretical advances in PCE with practical applications to real-world systems.

The study is organized as follows. Section~\ref{sec:methodology} details the intrusive time-dependent PCE methodology. Section~\ref{sec:results} presents numerical validations and comparisons between PCE and Monte Carlo (MC) simulations. Section~\ref{sec:conclusion} concludes and discusses future research directions.
\section{METHODOLOGY}
\label{sec:methodology}
\begin{figure}
\centering
    \captionsetup{width=\textwidth}
	\includegraphics[width=0.12\textwidth]{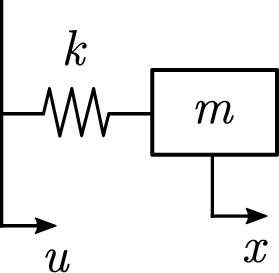}
	\caption{Spring-mass system with input $u$.}
 \label{fig:1_Spring_Mass}
\end{figure}

This section describes the PCE and TDF design strategies. Furthermore, the spring-mass system with its uncertainty is explained. Fig~\ref{fig:1_Spring_Mass} illustrates a single spring-mass system with an input $u$.
The equation of motion of the spring-mass system can be written as:\
\begin{subequations}
    \begin{align}
        m\ddot{x} + kx = ku\\
        \iff \ddot{x} + \omega_n^2x = \omega_n^2u,
    \end{align}
    \label{eq:spring_mass_EOM}
\end{subequations}
where $x$ is the position of the mass $m$. In this work it is assumed that the spring stiffness $k$ is uncertain, which directly correlates to an uncertainty in the natural frequency by $\omega_n=\sqrt{k/m}$. The mass is assumed to be constant with $m=1$. The residual energy for this system is defined as:
\begin{align}
    V_{res} = \frac{m}{2}\dot{x}^2 + \frac{k}{2}(x_f-x)^2,\label{eq:res_energy_nominal}
\end{align}
where $x_f$ is the desired position of the mass.
%
%
%
\subsection{Polynomial Chaos Expansion}
Polynomial Chaos is a well-established approach for representing a stochastic output as a weighted combination of polynomial functions of the random variables as inputs.
A PCE has the form of~\cite{singh_polynomial_2010}:
\begin{align}
    F(t,\zeta)=\sum^{\infty}_{i=0}a_i(t)\Psi_i(\zeta), \label{eq:PCE}
\end{align}
where $F$ is the output variable, $a_i$ are the PC-coefficients and $\Psi_i$ are the basis polynomial functions. Moreover, $t$ is the deterministic independent variable and $\zeta$ is the random variable. The orthogonality condition of the basis functions is given by~\cite{wang_polynomial_2020}:
\begin{align}
    \left<\Psi_i(\zeta),\Psi_j(\zeta)\right> = \int_\Omega\Psi_i(\zeta)\Psi_j(\zeta)f(\zeta)d\zeta = g_i^2\delta_{ij}, \label{eq:PCE_orthogonality}
\end{align}
where $g_i^2$ are the coefficients resulting from the evaluation of the inner product, and $\delta_{ij}$ is the Kronecker delta. $f(\zeta)$ is the pdf of $\zeta$. The Wiener–Askey scheme~\cite{xiu_wiener--askey_2002} provides the appropriate choice of polynomial basis functions for random variables with standard well-known distributions, as illustrated in Table~\ref{table:Wiener_Askey_scheme}.
\begin{table}
\centering 
\caption{Wiener-Askey scheme for polynomial chaos expansion. $k_l$ and $k_u$ are the user-defined lower and upper bounds.}
\label{table:Wiener_Askey_scheme}
\small
\begin{tabular}{c c c}
\toprule 
Distribution of $\zeta$ & Polynomial family & Support \\
\midrule
Gaussian & Hermite & $\left(-\infty,\infty\right)$\\
gamma & Laguerre & $\left[0,\infty\right)$\\
beta & Jacobi & $\left[k_l,k_u\right]$\\
uniform & Legendre & $\left[k_l,k_u\right]$\\
\bottomrule
\end{tabular}
\end{table}
The PC-coefficients can be determined using an intrusive or a non-intrusive approach. The intrusive approach requires the evaluation of inner products and may not yield closed-form expressions for certain nonlinear functions~\cite{saha_uncertainty_2021}. The non-intrusive approach can be considered a black-box approach, which requires commonly solving a least-squares problem to determine the coefficients \cite{wang_polynomial_2020}.
In this work, the intrusive approach is used, which means that the coefficients are determined by exploiting Galerkin projections. Consider~\eqref{eq:PCE} which can be approximated as:
\begin{align}
    F(t,\zeta)\approx F_{PC}=\sum^P_{i=0}a_i(t) \Psi_i(\zeta). \label{eq:PCE_v2}
\end{align}
The left-hand side of~\eqref{eq:PCE_v2} represents the known equation that need to be approximated by a series of $P$ orthogonal polynomials with unknown coefficients $a_i$, as shown on the right-hand side of~\eqref{eq:PCE_v2}. Applying Galerkin's method corresponds to taking the inner product of both sides of~\eqref{eq:PCE_v2} with $\Psi_j(\zeta)$. In combination with the orthogonality condition, this results in: 
\begin{align}
    \int F(t,\zeta) \Psi_j(\zeta) f(\zeta)d\zeta = g_j^2 a_j(t).\label{eq:PCE_galerkin_j}
\end{align}
Eqn. \eqref{eq:PCE_galerkin_j} can be solved to determine $a_j$. For problems with multiple random variables, the basis functions are determined by the tensor product of the one-dimensional basis functions. In general, a PCE with $d$-uncertainties takes the form
\begin{subequations}\label{eq:7}
  \begin{align}
        F(t,\zeta_1, \ldots,\zeta_d) = \sum\limits_{i=0}^{K}a_i(t)\Psi_i(\zeta_1, \ldots, \zeta_d), \\
        \Psi_i(\zeta_1, \ldots, \zeta_d) = \prod\limits_{l = 1}^d\Psi_{i_l}(\zeta_l), \label{eqn7b}\\
        0\leq i_l\leq P;\:\: \sum\limits_{l=1}^d i_l\leq P,
    \end{align}
\end{subequations}
and $K =\frac{(P+d)!}{P!d!}-1$ is the total number of coefficients \cite{wang_polynomial_2020}.
\subsection{Time-Delay Filter Designs} 
\label{sec:tdf}
In this section two different designs of the TDFs are introduced, namely the non-robust and robust TDF.
%
%
%
\subsubsection{Non-robust and Robust}
The main goal of applying a TDF is to place a pair of zeros to the set of underdamped poles, which lie at $\omega_d=\omega_n\sqrt{1-\zeta^2}$. For simplicity no damping ratio $\zeta$ is assumed to be zero in this work. Singh~\cite{singh_optimal_2009} defined an input shaper with one delay as:
\begin{subequations}
    \begin{align}
         G_{nonrob}(s) = A_0 + \left(A_1 e^{-s T}\right), \label{eq:G_nonrob}\\
        A_0 = \frac{e^{\left(\frac{\xi\pi}{\sqrt{1-\xi^2}}\right)}}{1+e^{\left(\frac{\xi\pi}{\sqrt{1-\xi^2}}\right)}};\:\: A_1 = 1-A_0;\:\: T = \frac{\pi}{\omega_d},
    \end{align}
\end{subequations}
which places one pair of zeros at the location of the underdamped poles. This is referred to as the non-robust TDF. $A_0$ and $A_1$ are the magnitudes of the proportional signal and $T$ is the delay time. The robust input shaper is selected to place two sets of zeros at the nominal location of the underdamped poles of the system resulting in the robust TDF:
\begin{align}
    G_{rob}(s) = \left(A_0 + A_1 e^{-s T}\right)^2. \label{eq:G_rob}
\end{align}
%
%
%
\subsubsection{Global Sensitivity Analysis}
A GSA TDF takes the form ~\cite{singh_optimal_2009}:
\begin{align}
    G(s)=\mathcal{A}_0 + \sum\limits_{i=1}^{N}\mathcal{A}_ie^{-s\mathcal{T}_i},
\end{align}
where $N$ is defined by the underlying TDF design ($N=1$ for non-robust, and $N \geq 2$ for robust). To determine the GSA TDF parameters, the following minimization problem is solved:
\begin{subequations} \label{eq:GSA_min_prob}
\begin{align}
    \min_{\mathcal{A}_i,\mathcal{T}_i} & J(\mathcal{A}_i,\mathcal{T}_i) = \Xi,\\
     \text{s.t.} & \sum\limits_{i=0}^{N}\mathcal{A}_i=1,\text{ } 0\leq\mathcal{A}_i,\label{eq:GSA_cons_1}\\
    & 0<\mathcal{T}_i<\mathcal{T}_{N} = t_f, \label{eq:GSA_cons_2}
\end{align}
\end{subequations}
where $J$ is selected based on the specific application $\Xi$ of the GSA TDF and the final time $t_f$ is set by the corresponding closed-form expression of the non-robust or robust TDF design for fair comparison.
\begin{figure}
\centering
    \captionsetup{width=1\linewidth}
	\includegraphics[width=0.4\textwidth]{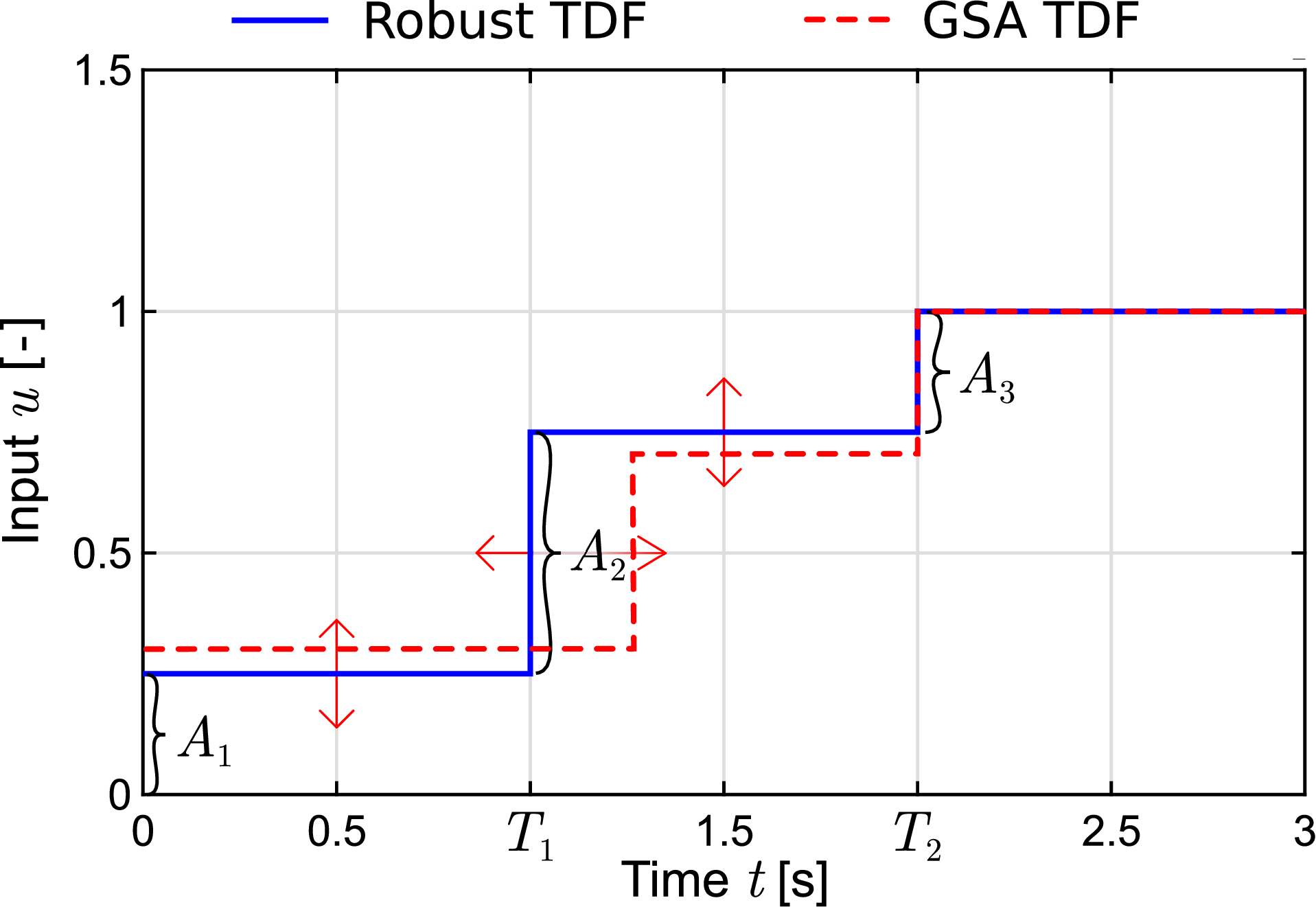}
	\caption{Functionality of a GSA TDF over time.}
 \label{fig:temp_05_GSA}
\end{figure}
\subsection{Spring-Mass System with Interval Uncertainty}
In this work, polynomial chaos is applied to a spring-mass system (see~\eqref{eq:spring_mass_EOM}) with uncertain natural frequency. To make the natural frequency uncertain, let $\omega_n \sim \mathcal{U}(\omega_{n,lb},\omega_{m,ub})$ be a uniform random variable where $(.)_{lb}$ and $(.)_{ub}$ are the lower and upper bound, respectively. Furthermore, the goal of this work is to account for time-varying uncertainties. This is represented by assuming that~\eqref{eq:spring_mass_EOM} holds in the interval $[0,t_1]$ and in $[t_1,t_2]$ the system is described by
\begin{align}
    \ddot{x} + \omega_m^2x = \omega_m^2u,\label{eq:2nd_order_system_general_t2}
\end{align}
where $\omega_m\sim \mathcal{U}(\omega_{m,lb},\omega_{m,ub})$ now comes from a second uniform distribution. In the context of this work, it is assumed that $\mathbb{E}[\omega_n] = \mathbb{E}[\omega_m] \mathrel{=:} \mu_{\omega}$ but $Var(\omega_n) \neq Var(\omega_m)$.\\
After rewriting both equations, the result is:
\begin{subequations}
    \begin{align}
        \ddot{x} = -\omega_n^2x + \omega_n^2u, \label{eq:2nd_order_system_general_sep_1}\\
        \ddot{x} = -\omega_m^2x + \omega_m^2u,\label{eq:2nd_order_system_general_sep_2}
    \end{align} \label{eq:2nd_order_system_general_combined}
\end{subequations}
where $\ddot{x}$ can now be seen as a function of $\omega_n$ and $\omega_m$, respectively. This allows to approximate $\ddot{x}$ with Polynomial Chaos where 
\begin{subequations}
    \begin{align}
        \ddot{x}(t,\omega_n) &= \ddot{x}(t,\zeta_1) \approx \sum^P_{i_1=0}c_{i_1}(t) \Psi_{i_1}(\zeta_1),\\ 
        \ddot{x}(t,\omega_m) &= \ddot{x}(t,\zeta_1, \zeta_2) \approx \sum^{K}_{i_2=0}c_{i_2}(t)\Psi_{i_2}(\zeta_1, \zeta_2), \label{eq:PCE_LHS_sm}
    \end{align}
\end{subequations}
and
\begin{subequations}
    \begin{align}
        \omega_n = \mu_{\omega} + \frac{\omega_{n,ub}-\omega_{n,lb}}{2}\zeta_1,\\
        \omega_m = \mu_{\omega} + \frac{\omega_{m,ub}-\omega_{m,lb}}{2}\zeta_2,\label{eq:zeta_dist}
    \end{align}
\end{subequations}
with $\zeta_1,\zeta_2\sim \mathcal{U}(-1,1)$. $\Psi_{i_1}$ and $\Psi_{i_2}$ are chosen as the Legendre polynomials because $\zeta_1, \zeta_2$ are uniformly distributed~\cite{xiu_wiener--askey_2002}.

Since the natural frequency is uncertain in the first time interval $[0,t_1]$, $x$ is also uncertain at $t_1$, which in turn serves as the initial condition for the second time interval $[t_1,t_2]$. Hence, for $t \geq t_1$, $x(t)$ depends not only on $\omega_m$ (and thus on $\zeta_2$), but also on $\zeta_1$. The right-hand sides of~\eqref{eq:2nd_order_system_general_combined} are similarly approximated by:
\begin{subequations}
    \begin{flalign}
         &-\omega_n^2 x(t,\zeta_1) + \omega_n^2 u(t) \approx \nonumber\\
         &-\left(\mu_{\omega} + \frac{\omega_{n,ub}-\omega_{n,lb}}{2}\zeta_1\right)^2\sum^P_{i_1=0}a_{i_1}(t)\Psi_{i_1}(\zeta_1)\nonumber\\
         &+ \left(\mu_{\omega} + \frac{\omega_{n,ub}-\omega_{n,lb}}{2}\zeta_1\right)^2 u(t),&& \\ 
        &-\omega_m^2 x (t,\zeta_1,\zeta_2) + \omega_m^2 u(t) \approx \nonumber\\
        &-\left(\mu_{\omega} + \frac{\omega_{m,ub}-\omega_{m,lb}}{2}\zeta_2\right)^2\sum^{K}_{i_2=0}a_{i_2} (t)\Psi_{i_2}(\zeta_1, \zeta_2) \label{eq:PCE_RHS_sm} \nonumber \\
        &+ \left(\mu_{\omega} + \frac{\omega_{m,ub}-\omega_{m,lb}}{2}\zeta_2\right)^2 u(t), &&
    \end{flalign}
\end{subequations}
where $\ddot{a}_{i_k} = c_{i_k}$, $k\in\{1,2\}$ . Now Galerkin's method can be applied to obtain a closed-form expression for $c_{i_k}$ in terms of $a_{i_k}$ and $u$. To solve this system of differential equations, the MATLAB function \texttt{ode45} is used first for $k=1$ (i.e., over the interval $[0,t_1]$) and then for $k=2$ (i.e., over the interval $[t_1,t_2]$). The solution derived from \texttt{ode45} provides values for both $a_{i_k}$ and $b_{i_k} = \dot{a}_{i_k}$ at each timestamp.

To ensure continuity of $x$ and $\dot{x}$ at $t_1$, the following continuity conditions must be satisfied:
\begin{subequations}\label{eq:cont_cons}
    \begin{align}
        \sum^P_{i_1=0}a_{i_1}(t_1)\Psi_{i_1}(\zeta_1) = \sum^{K}_{i_2=0}a_{i_2}(t_1)\Psi_{i_2}(\zeta_1, \zeta_2), \label{eq:is_cont_1}\\
        \sum^P_{i_1=0}b_{i_1}(t_1)\Psi_{i_1}(\zeta_1)= \sum^{K}_{i_2=0}b_{i_2}(t_1)\Psi_{i_2}(\zeta_1, \zeta_2). \label{eq:is_cont_2}
    \end{align} \label{eq:is_cont_1and2}
\end{subequations}
As it has been shown in~\eqref{eqn7b}, both $a_{i_2}$ and $b_{i_2}$ are coefficients of polynomials consisting of $\zeta_1$ and $\zeta_2$. Hence, $a_{i_2}$ and $b_{i_2}$ can be rewritten using a different notation as follows:
\begin{subequations}
    \begin{align}
        \sum^{K}_{i_2=0}a_{i_2}(t)\Psi_{i_2}(\zeta_1, \zeta_2) = \sum^{P_2}_{j_{2} = 0} \sum^{P_1}_{j_{1} = 0} a_{j_1 j_2}(t) \Psi_{j_1}(\zeta_1) \Psi_{j_2}(\zeta_2) ,\\
            \sum^{K}_{i_2=0}b_{i_2}(t)\Psi_{i_2}(\zeta_1, \zeta_2) = \sum^{P_2}_{j_{2} = 0} \sum^{P_1}_{j_{1} = 0} b_{j_1 j_2} (t) \Psi_{j_1}(\zeta_1) \Psi_{j_2}(\zeta_2),          
    \end{align} \label{eq:18}
\end{subequations}
where $\left(P_1 + 1\right)\times \left(P_2 + 1\right) = K + 1$. By applying Galerkin's projection method to~\eqref{eq:is_cont_1and2}, together with~\eqref{eq:18}, the values of $a_{i_2}$ and $b_{i_2}$ at $t_1$ can be obtained, which are essential for initializing \texttt{ode45} in the second time interval $[t_1,t_2]$. First,
\begin{subequations}
\label{eq:galerkin_ai2}
\begin{flalign}
     &\sum^P_{i_1=0}a_{i_1}(t_1)\Psi_{i_1}(\zeta_1) = \sum^{P_2}_{j_{2} = 0} \sum^{P_1}_{j_{1} = 0} a_{j_1 j_2}(t) \Psi_{j_1}(\zeta_1) \Psi_{j_2}(\zeta_2), \label{eq:continuity_t2}
\end{flalign}
where the inner product is taken as:
\begin{flalign}
     &\implies \int\limits_{-1}^{1}\int\limits_{-1}^{1}\biggl(\sum_{i_1=0}^P a_{i_1}(t_1)\Psi_{i_1}(\zeta_1)\biggr)\Psi_{w_1}(\zeta_1)\Psi_{w_2}(\zeta_2) \nonumber\\
     &\phantom{\implies \int_{-1}^{1}\int_{-1}^{1}} f(\zeta_1)f(\zeta_2)\,d\zeta_1 d\zeta_2 =\nonumber\\
     &\phantom{\implies}\int\limits_{-1}^{1}\int\limits_{-1}^{1}\biggl(\sum^{P_2}_{j_{2} = 0} \sum^{P_1}_{j_{1} = 0} a_{j_1 j_2}(t) \Psi_{j_1}(\zeta_1) \Psi_{j_2}(\zeta_2)\biggr) \nonumber\\
     &\phantom{\implies \int_{-1}^{1}\int_{-1}^{1}}\Psi_{w_1}(\zeta_1)\Psi_{w_2}(\zeta_2)f(\zeta_1)f(\zeta_2)d\zeta_1d\zeta_2,&&
\end{flalign}
where the separation of $\zeta_1$ and $\zeta_2$ dependencies leads to:
\begin{flalign}
     &\implies \int\limits_{-1}^{1}\Psi_{w_2}(\zeta_2)f(\zeta_2)\int\limits_{-1}^{1}\biggl(\sum_{i_1=0}^P a_{i_1}(t_1)\Psi_{i_1}(\zeta_1) \Psi_{w_1}(\zeta_1) \nonumber\\
     &\phantom{\implies \int_{-1}^{1}\int_{-1}^{1}} f(\zeta_1) \biggr)\,d\zeta_1 d\zeta_2 =\nonumber\\
     &\phantom{\implies}\int\limits_{-1}^{1}\biggl( \sum^{P_2}_{j_{2} = 0} \Psi_{j_2}(\zeta_2)\Psi_{w_2}(\zeta_2)f(\zeta_2)\biggr)\int\limits_{-1}^{1}\biggl( \sum^{P_1}_{j_{1} = 0} a_{j_1 j_2}(t)  \nonumber\\
     &\phantom{\implies \int_{-1}^{1}\int_{-1}^{1}}\Psi_{j_1}(\zeta_1)\Psi_{w_1}(\zeta_1)f(\zeta_1)\biggr)d\zeta_1d\zeta_2,&&
\end{flalign}
with further simplifications:
\begin{flalign}
     \iff & \frac{a_{w_1}(t_1)}{2w_1 +1}\int\limits_{-1}^{1}\Psi_{w_{2}} f(\zeta_2) d\zeta_2 = \frac{a_{w_1 w_2}(t_1)}{(2w_{2}+1)(2w_{1}+1)},&&
\end{flalign}
\begin{flalign}
     \iff & a_{w_1}(t_1)\int\limits_{-1}^{1}\Psi_{w_{2}}(\zeta_2)\cdot1 f(\zeta_2) d\zeta_2 
     = \frac{a_{w_1 w_2}(t_1)}{2w_{2}+1}, &&
\end{flalign}
\begin{flalign}
     \iff & a_{w_1}(t_1)\int\limits_{-1}^{1}\Psi_{w_{2}}(\zeta_2)\Psi_0(\zeta_2)f(\zeta_2) d\zeta_2 = \frac{a_{w_1 w_2}(t_1)}{2w_{2}+1} &&
\end{flalign}
\begin{flalign}
     \iff & a_{w_1}(t_1) \frac{\delta_{w_2 0}}{2w_{2}+1} = \frac{a_{w_1 w_2}(t_1)}{2w_{2}+1}, &&
\end{flalign}
\begin{flalign}
     \iff & a_{w_1}(t_1) \delta_{w_2 0} = a_{w_1 w_2}(t_1). &&
\end{flalign}
This allows to write the PCE coefficients as:
\begin{flalign}
     \implies & a_{w_1 w_2}(t_1)=
     \begin{cases}
         a_{w_1}(t_1), &w_2=0\\
         0, &w_2>0.
     \end{cases}
     &&
\end{flalign} \label{eq:19}
\end{subequations}

\noindent
In particular,~\eqref{eq:19} leads to the following conclusion:
    \begin{align}
       a_{i_2}(t_1) := a_{j_1 j_2}(t_1)=
        \begin{cases}
       a_{i_1}(t_1),& j_2 = 0\\
        0,& j_2 > 0,
        \end{cases}
    \end{align}
\noindent
In a similar manner, it can be shown that:
\begin{align}
b_{i_2}(t_1) := b_{j_1 j_2}(t_1)=
        \begin{cases}
        b_{i_1}(t_1),& j_2 = 0\\
        0,& j_2 > 0.
        \end{cases}
\end{align}
A TDF can now be applied to control $u$ over time. This is done by updating $u$ in each timestamp through
\begin{align}
    u_t=\left(A_0+\sum\limits_{i=1}^{N}A_i\cdot \mathcal{H}(t-T_i)\right)u, \label{eq:hvs_tdf}
\end{align}
where $\mathcal{H}$ is the Heaviside function. The implementation of the TDF for the spring-mass system aims to minimize the total vibration in the system at $t_2$. To compare different TDF designs the expected residual energy $V_{PC,res}$ with PCE is calculated as:
\begin{align}
    V_{PC,res} = &\text{ }0.5\left(\sum^{K}_{i=0}b_{2i}\Psi_i(\zeta_1,\zeta_2)\right)^2 \nonumber\\
    &+ 0.5\left(\sum^{K}_{i=0}a_{2i}\Psi_i(\zeta_1,\zeta_2)-1\right)^2. \label{eq:vres_pce}
\end{align}
Hence, $\Xi$ from~\eqref{eq:GSA_min_prob} takes the form of $\mathbb{E}[V_{res}]$ and $Var(V_{res})$.
%
%
%
\section{RESULTS}
\label{sec:results}
%
%
%
\subsection{Comparison of MC and PCE}
PCE serves as an appropriate, and in some cases superior, alternative to MC simulations. It must converge to results consistent with those obtained from MC. Additionally, both the time required to construct the PCE model and the subsequent computational time should be shorter than those associated with a MC simulation.
With respect to convergence, this study specifically examines the expected value and variance of $x$ and $\dot{x}$, since the primary objective is to minimize the expected value and variance of the residual energy as in~\eqref{eq:vres_pce}. In the following analysis, it is assumed that $\omega_n\sim \mathcal{U}(0.75\pi, 1.25\pi)$ and $\omega_m\sim \mathcal{U}(0.5\pi, 1.5\pi)$ are the sample distributions for the natural frequency, yielding $\mu_{\omega}=\pi$. These distributions are selected solely for demonstration purposes. Furthermore, $K$ is chosen to be $\frac{(P+2)!}{P!2!}-1$ similar to how $K$ is set in~\eqref{eq:7}. As a result, $P$ becomes the sole variable that determines the size of the PCE models in both time intervals. Finally, 
$t_1$ is set to 100 s and $t_2$
 to 200 s.
 
\begin{figure}
\centering
    \captionsetup{width=1\linewidth}
	\includegraphics[width=0.4\textwidth]{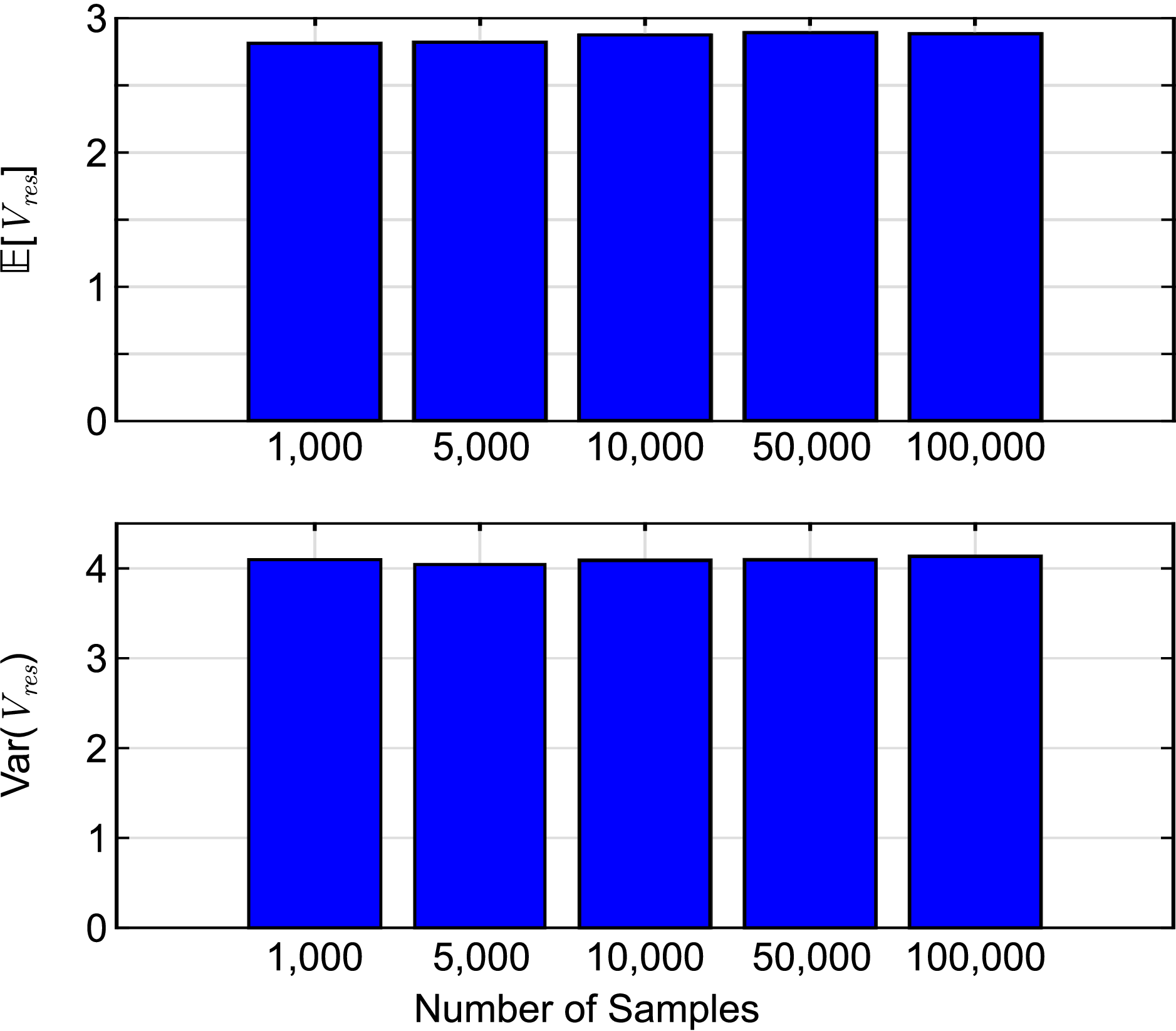}
	\caption{Expected residual energy and its variance for different MC sample sizes.}
 \label{fig:3_MC}
\end{figure}

\begin{figure}
\centering
    \captionsetup{width=1\linewidth}
	\includegraphics[width=0.4\textwidth]{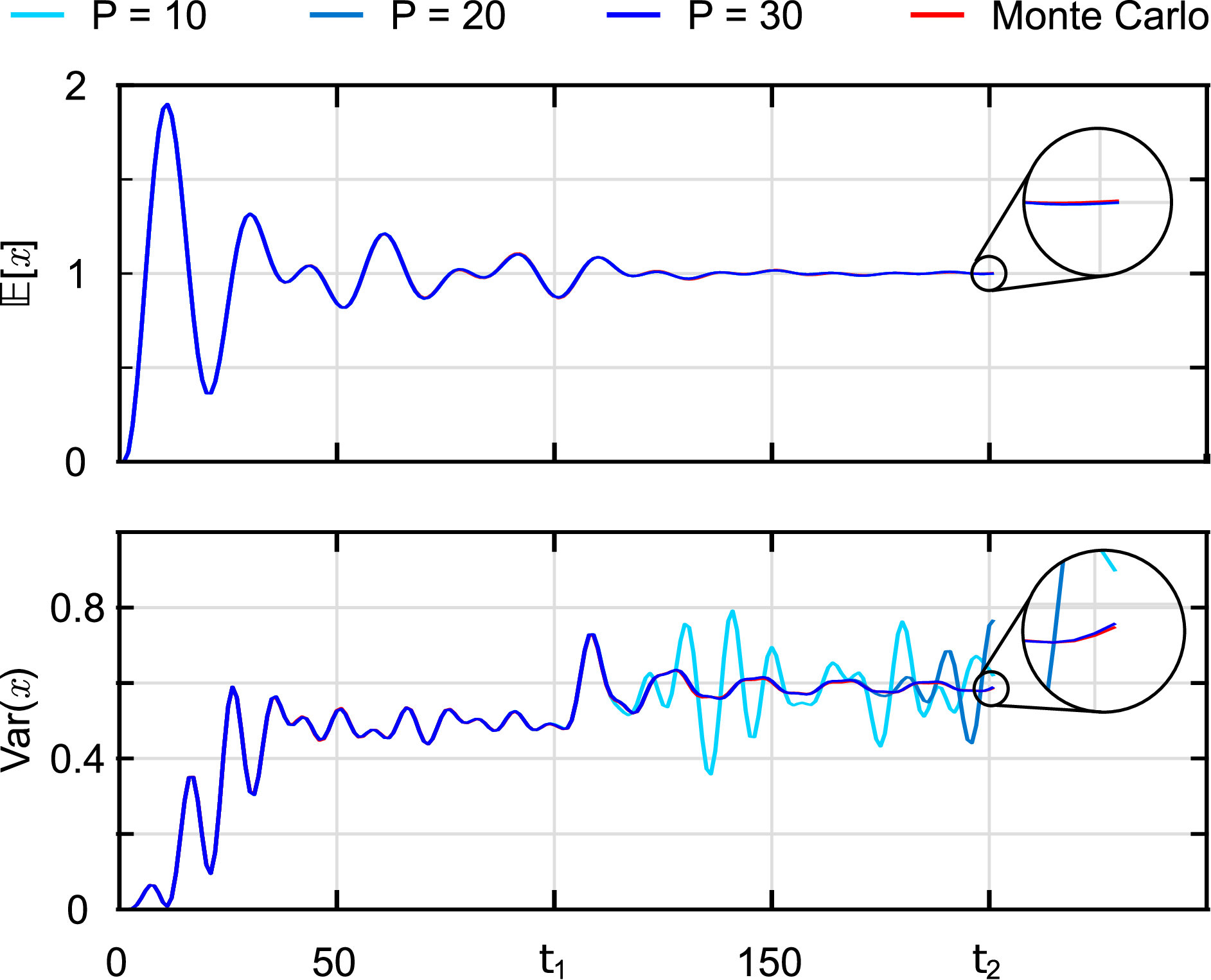}
	\caption{Convergence of PCE for expected value and variance for the position $x$.}
 \label{fig:5_Position_E_and_Var}
\end{figure}

\begin{figure}
\centering
    \captionsetup{width=1\linewidth}
	\includegraphics[width=0.4\textwidth]{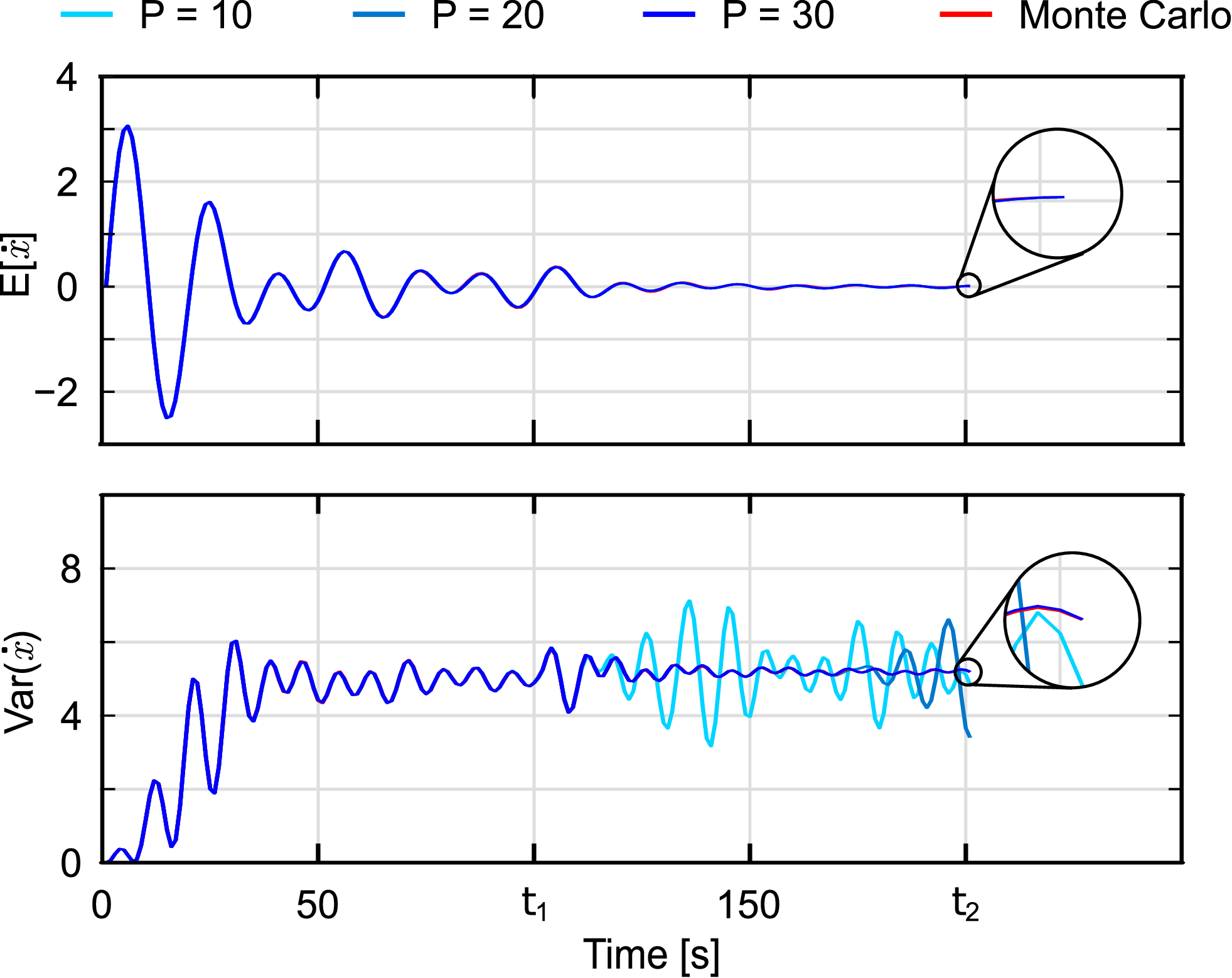}
	\caption{Convergence of PCE for expected value and variance for the velocity $\dot{x}$.}
 \label{fig:6_Velocity_E_and_Var}
\end{figure}

\begin{figure}
\centering
    \captionsetup{width=1\linewidth}
	\includegraphics[width=0.4\textwidth]{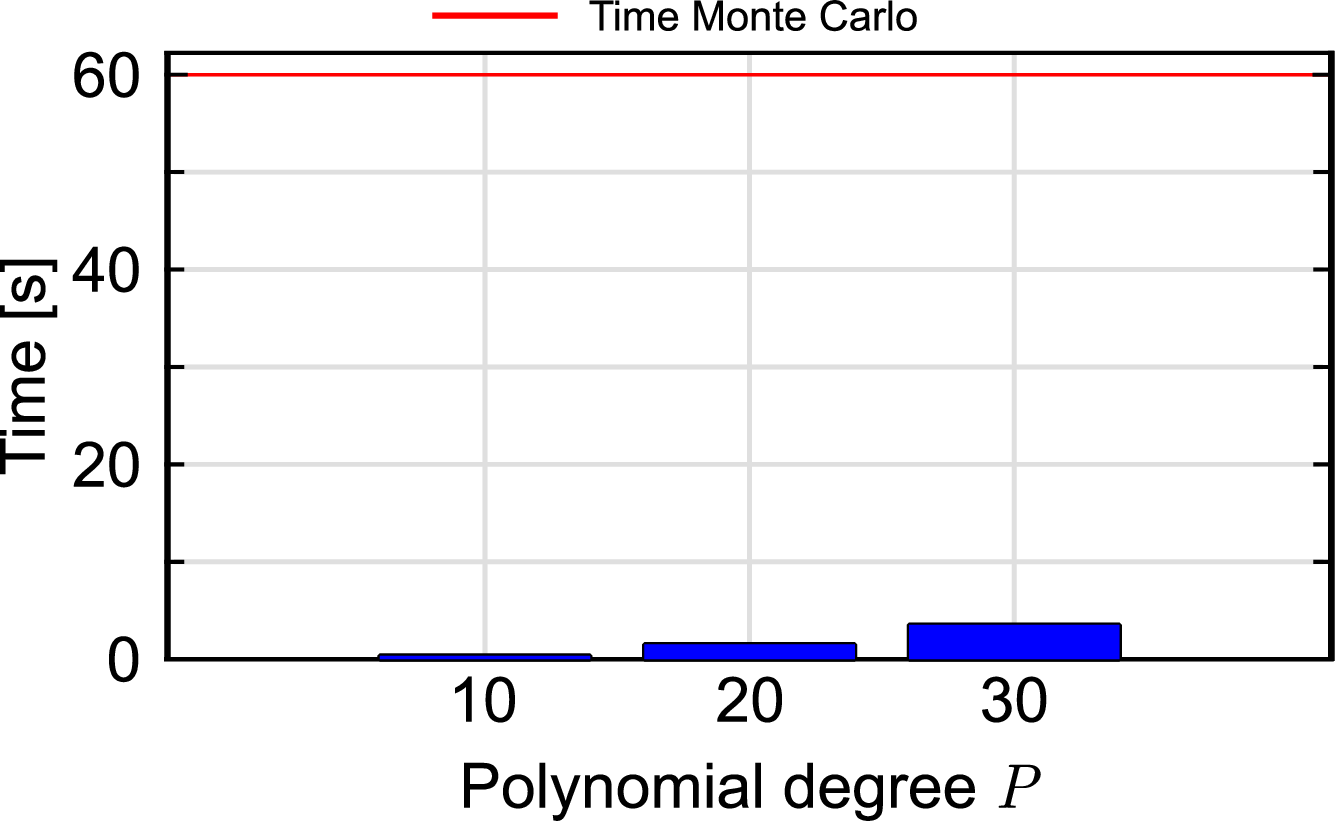}
	\caption{PCE computational time for different degrees.}
 \label{fig:4_PCE_time}
\end{figure}

Given both distributions, a MC simulation is conducted with different sample sizes ranging from $1,000$ samples to $100,000$ samples. Using pairs of $(\zeta_1, \zeta_2)$ taken from the distribution provides the necessary samples to calculate the expected residual energy and its variance for $t_2$. In Fig.~\ref{fig:3_MC} it can be seen that after $10,000$ samples, both measures converge. Therefore, the time comparison of MC to PCE is with respect to this amount of $10,000$.
Figs.~\ref{fig:5_Position_E_and_Var} and~\ref{fig:6_Velocity_E_and_Var} show the convergence of PCE for $x$ and $\dot{x}$ for different choices of $P$. For $x$ and $\dot{x}$, the expected value of PCE is found to quickly converge to MC with respect to $P$, and there is no significant difference between a PCE with degree $10$ and degree $30$. The variance converges much slower, with divergence from MC occurring up to degree $20$ in the second interval, though there is no noticeable difference between degree $10$ or $30$ in the first interval. The inset in both figures show an excellent performance of PCE with $P=30$ at time $t_2$. This finding motivates the choice of $P=30$ for further analysis. Additionally, Figs.~\ref{fig:5_Position_E_and_Var} and~\ref{fig:6_Velocity_E_and_Var} show that the method used to initialize the PCE coefficients at the beginning of the second interval as shown in~\eqref{eq:continuity_t2} ensures continuity in $t_1$.
In terms of computational complexity, the construction of both PCE polynomials combined (first and second interval) with $P=30$ is significantly faster than MC. Fig.~\ref{fig:4_PCE_time} shows that, regarding time, PCE outperforms a MC simulation with a sample size of $10,000$ by a factor of approximately $12$.
The above results are achieved on an Intel Core i7-4700KF, 3400Mhz with 32 GB RAM where the tolerance of \texttt{ode45} was set to $10^{-12}$ in terms of relative and absolute error. For this analysis $u=1, \forall t\in[0,t_2]$.

%
%
%
\subsection{Application of TDF}
The obtained PCE model is combined with each TDF design from Section ~\ref{sec:tdf}, with the robust TDF selected as the underlying design for the GSA TDF. The objective is to identify the TDF design that minimizes both the expected residual energy and the variance of $V_{res}$ at $t_2$. Since the underlying spring–mass system is assumed to be undamped, $\xi$ is set to 0 for the robust and non-robust TDF. for both the robust and non-robust TDF, yielding $A_0=A_1=0.5$. Furthermore, $\omega_d$ is set equal to $\mu_{\omega}$ resulting in $T_1=1$.\\
To determine the parameters of the GSA TDF, MATLAB’s \texttt{fmincon} function is employed. For case 1) \mbox{$\Xi=\Xi_1=\mathbb{E}\left[V_{PC,res}\right]$} and case 2) \mbox{$\Xi=\Xi_2=Var\left(V_{PC,res}\right)$}. The enforced constraints are those specified in~\eqref{eq:GSA_min_prob}, and the robust TDF parameters serve as the initial values for the optimizer. As a result, the two sets of parameters for the GSA TDF differ, and the choice between minimizing the expected value or the variance ultimately depends on the intended application.\\
Table~\ref{tab:res_en_TDF} shows that all TDF designs significantly reduce the expected residual energy and its variance, with the variance reduced by at least a factor of 20 by all three TDFs. In general, the GSA TDFs give the lowest values outperforming the robust TDF by a factor of 2 and 5 in terms of expected value and variance, respectively. Optimization with regard to the variance results in a nearly constant residual energy for all pairs of $(\omega_n,\omega_m)$. This implies that the TDF GSA ensures that the desired performance remains effectively insensitive to variations arising from the uncertainty in the system. Table~\ref{table:magnitudes_and_switch_times_TDFs} lists the corresponding switch profiles for the TDF designs. Interestingly, \texttt{fmincon} keeps $\mathcal{T}_1$ as initialized and only slightly adjusts $\mathcal{A}_1,\mathcal{A}_2$ and $\mathcal{A}_3$ where both cases result in higher values for $\mathcal{A}_1$ and $\mathcal{A}_3$
and a lower value for $\mathcal{A}_2$. It should be noticed that different initial values for \texttt{fmincon} might give deviating parameters for the GSA TDF since these are suboptimal solutions.
\begin{table}
\begin{center}
    \setlength{\tabcolsep}{4pt}
    \begin{tabular}{lccccc}
        \cmidrule(lr){2-6}
         & \vtop{\hbox{\strut $u=1$}\hbox{\strut $\forall t$}} & \vtop{\hbox{\strut Non-robust}\hbox{\strut TDF}} & \vtop{\hbox{\strut Robust}\hbox{\strut TDF}} & \vtop{\hbox{\strut GSA}\hbox{\strut TDF}\hbox{\strut $\Xi_1$}} & \vtop{\hbox{\strut GSA}\hbox{\strut TDF}\hbox{\strut $\Xi_2$}}\\
         \midrule
        $\scriptsize\mathbb{E}[V_{res}]$ & 2.8899 & 0.1453 & 0.0129 & 0.0062 & 0.0068\\
        $\scriptsize Var(V_{res})$ & 4.1211 & 0.0358 & 0.0006 & 0.00009 & 0.00006\\
    \end{tabular}
    \caption{Expected value and variance of residual energy for non-robust, robust and GSA TDF}
    \label{tab:res_en_TDF}
\end{center}    
\end{table}
\begin{table}
\begin{center}
\begin{tabular}{cccccc}
 & \vtop{\hbox{\strut Non-robust}\hbox{\strut TDF}} & \vtop{\hbox{\strut Robust}\hbox{\strut TDF}} & \vtop{\hbox{\strut GSA TDF}\hbox{\strut $\Xi_1$}} & \vtop{\hbox{\strut GSA TDF}\hbox{\strut $\Xi_2$}}\\
\midrule
\addlinespace[3pt]
$T_1/\mathcal{T}_1$ & 1 & 1 & 1 & 1\\
$T_2/\mathcal{T}_2$ & - & 2 & 2 & 2\\
\hline
\addlinespace[3pt]
$A_0/\mathcal{A}_1$ & 0.5 & 0.25 & 0.2617 & 0.2673\\
$A_1/\mathcal{A}_2$ & 0.5 & 0.5 & 0.4745 & 0.4673\\
$A_2/\mathcal{A}_3$ & - & 0.25 & 0.2638 & 0.2654\\
\hline
\end{tabular}
\caption{Magnitudes and switch times for non-robust, robust and GSA TDF}
\label{table:magnitudes_and_switch_times_TDFs}
\end{center}
\end{table}
Although the GSA TDF ensures the lowest expected residual energy and variance, there exist pairs of $(\omega_n,\omega_m)$ where the residual energy in $t_2$ is lower when using a robust TDF. 
\begin{figure}
\centering
    \captionsetup{width=1\linewidth}
	\includegraphics[width=0.4\textwidth]{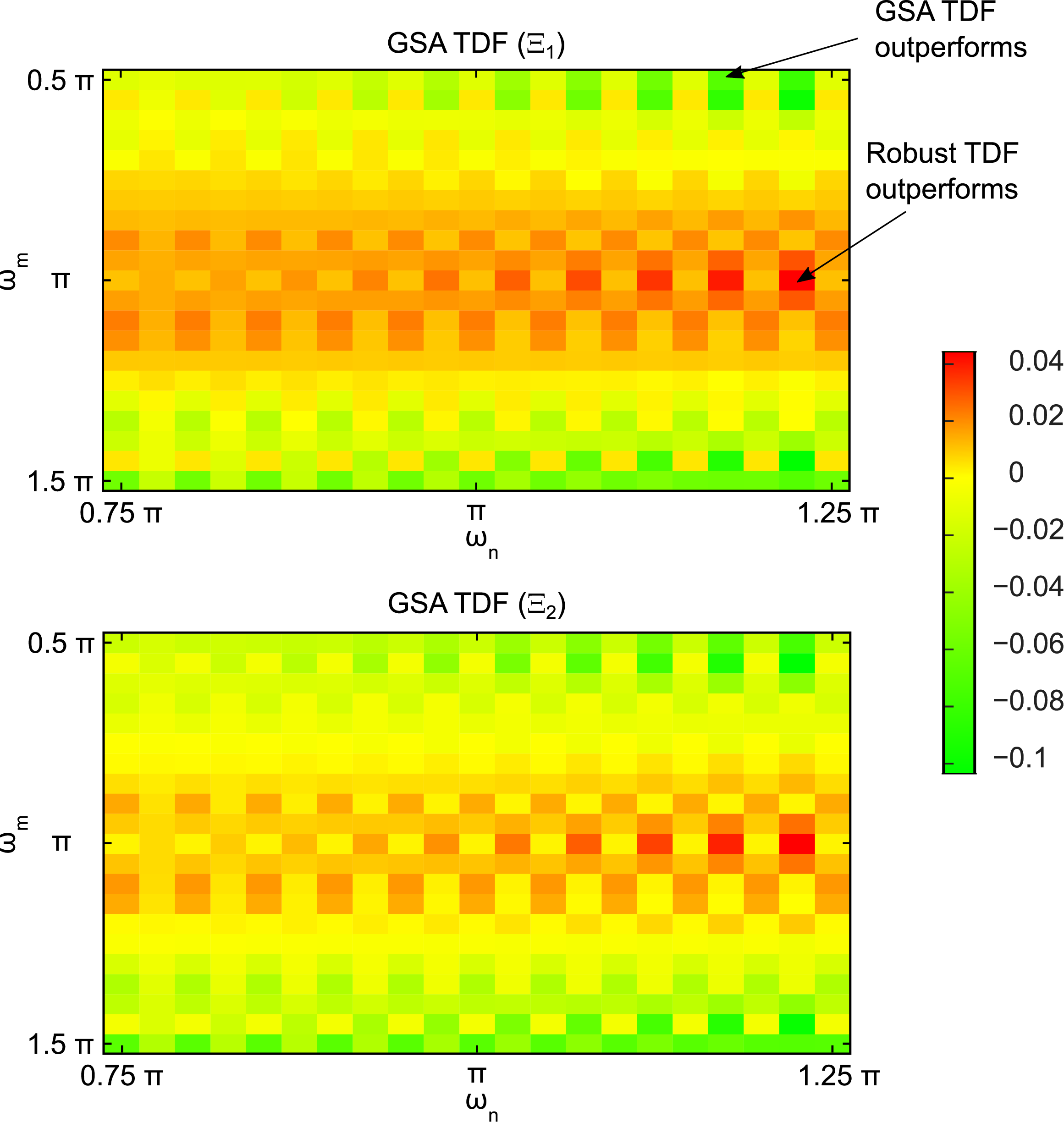}
	\caption{Differences between residual energy of robust and GSA TDF for different pairs of $(\omega_n,\omega_m)$. First heatmap shows the GSA TDF ($\Xi_1$) with minimal expected residual energy and the second the GSA TDF ($\Xi_2$) with minimal variance.}
 \label{fig:7_Heatmap}
\end{figure}
Fig.~\ref{fig:7_Heatmap} shows the combinations of $\omega_n$ and $\omega_m$ for which this is the case. Notice that if $\omega_m$ is close to its mean, the robust TDF outperforms the GSA TDF regardless of the value of $\omega_n$. However, the figure also shows that the actual difference between the residual energy of the robust and GSA TDF is small, especially compared to the borderline cases of $(\omega_n,\omega_m)$ for which the GSA TDF ensures significantly lower values. When both GSA TDFs are compared directly, the described behavior is more apparent for the GSA TDF that was designed with $\Xi_2$.
%
%
%
\section{CONCLUSIONS}
\label{sec:conclusion}
This work has demonstrated the efficacy of applying intrusive PCE to a spring mass system with uncertain and time-varying natural frequency. Compared to Monte Carlo with an appropriate number of samples, PCE converges to the same expected value and variance, while also providing a substantial reduction in computational time. Due to the computational intensity associated with the large number of samples required by the Monte Carlo method, PCE provides a computationally inexpensive alternative for determining the appropriate parameters of the time-delay filter, even in the presence of interval-dependent uncertainty. By employing PCE as a framework for designing input shaping via a time-delay filter, the residual energy of the system can be reduced with greater efficiency and reduced computational effort. Considering the computational burden of Monte Carlo simulations, this approach is particularly advantageous for vibration suppression and residual energy reduction in nonlinear systems. Future work will extend these techniques to nonlinear and higher-fidelity models. The intrusive PCE method employed in this study provides a greater analytical insight into both the system response and the objective function. Nevertheless, future work will explore non-intrusive PCE methods, which offer broader versatility and applicability to complex and nonlinear systems. Finally, experimental validation will be performed to demonstrate the practical implementation of PCE in conjunction with input shaping for vibration mitigation in aerial payload transport.



\bibliographystyle{IEEEtran}
\bibliography{references}
\addtolength{\textheight}{-12cm}
\end{document}